\documentclass[conference]{IEEEtran}
\usepackage{cite}
\usepackage{amsmath,amssymb,amsfonts}
\usepackage{algorithmic}
\usepackage{graphicx}
\usepackage{textcomp}
\usepackage{xcolor}

\usepackage[frozencache,cachedir=.]{minted}
    \setminted{fontsize=\footnotesize}
    \definecolor{mygray}{gray}{0.95}
\usepackage{url}

\usepackage[hidelinks]{hyperref}
\usepackage[nameinlink]{cleveref}

\newcommand{\codeword}[1]{\texttt{\textbf{\footnotesize#1}}}
\newcommand{\tool}{\textsc{DOCER\_tool}}

\def\BibTeX{{\rm B\kern-.05em{\sc i\kern-.025em b}\kern-.08em
    T\kern-.1667em\lower.7ex\hbox{E}\kern-.125emX}}
\begin{document}

\title{Wait, wasn't that code here before?\\Detecting Outdated Software Documentation}

\author{\IEEEauthorblockN{Wen Siang Tan}
\IEEEauthorblockA{\textit{School of Computer Science} \\
\textit{University of Adelaide}\\
Adelaide, SA, Australia \\
wensiang.tan@adelaide.edu.au}
\and
\IEEEauthorblockN{Markus Wagner}
\IEEEauthorblockA{\textit{Department of Data Science \& AI} \\
\textit{Monash University}\\
Melbourne, VIC, Australia \\
markus.wagner@monash.edu}
\and
\IEEEauthorblockN{Christoph Treude}
\IEEEauthorblockA{\textit{School of Computing and Information Systems} \\
\textit{The University of Melbourne}\\
Melbourne, VIC, Australia \\
christoph.treude@unimelb.edu.au}
}

\maketitle

\begin{abstract}
Encountering outdated documentation is not a rare occurrence for developers and users in the software engineering community. To ensure that software documentation is up-to-date, developers often have to manually check whether the documentation needs to be updated whenever changes are made to the source code. In our previous work, we proposed an approach to automatically detect outdated code element references in software repositories and found that more than a quarter of the 1000 most popular projects on GitHub contained at least one outdated reference. In this paper, we present a GitHub Actions tool that builds on our previous work's approach that GitHub developers can configure to automatically scan for outdated code element references in their GitHub project's documentation whenever a pull request is submitted.

Video—\url{https://www.youtube.com/watch?v=4cA10vdlmns}
\end{abstract}


\section{Introduction}
\begin{figure}[htbp]
    \centering
    \includegraphics[width=0.49\textwidth]{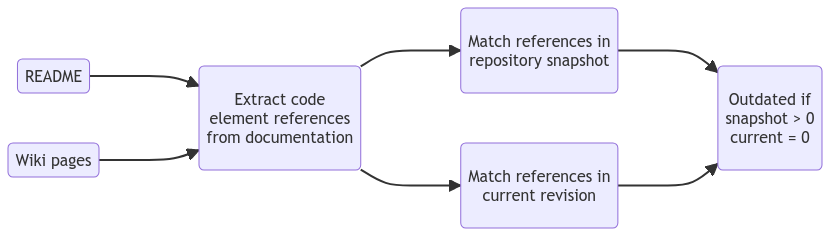}
    \caption{Overview of the DOCER approach introduced in our previous paper}
    \label{fig:docer_overview}
\end{figure}

Not only developers but also users often find encountering outdated software documentation a frustrating experience. In our previous work~\cite{tan2022detecting}, we found that 28.9\% of the top 1000 most popular projects\footnote{Top 1000 projects ranked by the number of stars} on GitHub contain at least one outdated reference to source code in their documentation. In the same paper, we proposed an approach named DOCER (Detecting Outdated Code Element References) to automatically detect outdated code element references in software repository documentation. The approach works by extracting code element references from documentation (README and wiki pages) using a list of regular expressions. These extracted references include variables, functions and class names found in the documentation such as \codeword{HttpClient}, \codeword{Promise.reject(err)} and \codeword{ArrayList<String>}. To determine if a reference is outdated, we match the reference to two revisions of the source code: the repository snapshot when the documentation was last updated and the current revision. We compare the number of instances found in the two versions and flag the reference as outdated if it existed in the snapshot but is no longer found in the current revision. \Cref{fig:docer_overview} shows an overview of the DOCER approach.

In our previous paper, we provided an implementation that developers can use to scan for outdated code element references. However, running the script whenever new changes are proposed may be mundane and repetitive. To simplify this process, we created a tool based on GitHub Actions workflow that is automatically triggered whenever a pull request is submitted to the repository. This workflow automates all the steps mentioned above and reports outdated references by commenting on the pull request. 

In the following sections of this paper, we provide an in-depth introduction to the tool's implementation (Section~\ref{sec:tool}), and describe real-world examples where the DOCER approach successfully detected outdated documentation (Section~\ref{sec:examples}). Limitations of the tool are discussed in Section~\ref{sec:limitations} before we conclude the paper with related (Section~\ref{sec:related}) and future work (Section~\ref{sec:conclusion}).

\section{Tool}
\label{sec:tool}
In this section, we introduce: (1) the GitHub Actions workflow that the tool is based on, (2) an example repository showing how the tool can be configured to run whenever a pull request is submitted, and (3) how false positives reported by the tool can be ignored.

\subsection{Implementation}
\label{sec:tool_implementation}
GitHub Actions,\footnote{\url{https://github.com/features/actions}} a feature on GitHub, enables developers to automate workflows based on events. This feature is typically employed for building Continuous Integration and Continuous Delivery (CI/CD) pipelines. We created the tool using GitHub Actions because it provides developers a convenient way to integrate the tool with existing GitHub projects. Developers also have the flexibility to configure their projects in a way that the tool automatically scans for outdated code element references in their documentation, whenever a pull request is submitted.

The workflow is defined by a YAML file\footnote{\url{https://yaml.org/}} containing a series of actions that gets executed when the workflow is triggered. To begin, we list the name of the workflow (DOCER), the events that trigger the workflow (pull requests), followed by the name of the GitHub-hosted runner\footnote{\url{https://docs.github.com/en/actions/using-github-hosted-runners/about-github-hosted-runners}} (latest Long Term Support version of Ubuntu) and the permissions needed for the job (read repository contents and write to pull requests).
\begin{minted}[bgcolor=mygray]{yaml}
name: DOCER

on: pull_request

jobs:
  run:
    runs-on: ubuntu-latest
    permissions:
      contents: read
      pull-requests: write
    steps:
\end{minted}

The rest of the file defines the steps to execute in the workflow. Three repositories are cloned on the runner (repositories containing the source code, wiki pages, and scripts for the analysis) using a GitHub Action named checkout.\footnote{\url{https://github.com/actions/checkout}}
\begin{minted}[bgcolor=mygray]{yaml}
- name: Checkout repository
  uses: actions/checkout@v3
  with:
    repository: ${{ github.repository }}
    ref: ${{ github.event.pull_request.head.sha }}
    path: repo
    fetch-depth: 0

- name: Checkout wiki
  continue-on-error: true
  uses: actions/checkout@v3
  with:
    repository: ${{ github.repository }}.wiki
    path: wiki

- name: Checkout tool
  uses: actions/checkout@v3
  with:
    repository: wesleytanws/DOCER_tool
    path: tool
\end{minted}

Once the repositories are cloned, the runner possesses all the necessary files to scan for outdated references. The workflow then commences the analysis, installs the necessary Python packages used by the report, generates the report and finally stores the results in an environment variable.
\begin{minted}[bgcolor=mygray]{yaml}
- name: Run tool
  run: |
    bash tool/analysis.sh

    pip install pandas
    pip install numpy

    echo 'report<<EOF' >> $GITHUB_ENV
    python tool/report.py ${{ github.repository }} \
        ${{ github.run_id }} >> $GITHUB_ENV
    echo 'EOF' >> $GITHUB_ENV
\end{minted}

In the case where merging the pull request may result in outdated documentation, the workflow uses a GitHub Action named github-script\footnote{\url{https://github.com/actions/github-script}} to post a comment on the pull request listing the potentially outdated references.
\begin{minted}[bgcolor=mygray]{yaml}
- name: Comment on pull request
  if: ${{ env.report }}
  uses: actions/github-script@v6
  env:
    report: ${{ env.report }}
  with:
    script: |
      github.rest.issues.createComment({
        issue_number: context.issue.number,
        owner: context.repo.owner,
        repo: context.repo.repo,
        body: process.env.report
      })
\end{minted}

Figuring out why a code element reference has been flagged as potentially outdated can be challenging, especially when there are numerous modifications in the pull request. This final step uploads the report and summary files to GitHub using a GitHub Action named upload-artifact,\footnote{\url{https://github.com/actions/upload-artifact}} allowing developers to view the full report.
\begin{minted}[bgcolor=mygray]{yaml}
- name: Upload artifact
  if: ${{ env.report }}
  uses: actions/upload-artifact@v3
  with:
    name: report
    path: |
      output/report.csv
      output/summary.csv
      output/summary.md
\end{minted}

\begin{figure}[htbp]
    \centering
    \includegraphics[width=0.45\textwidth]{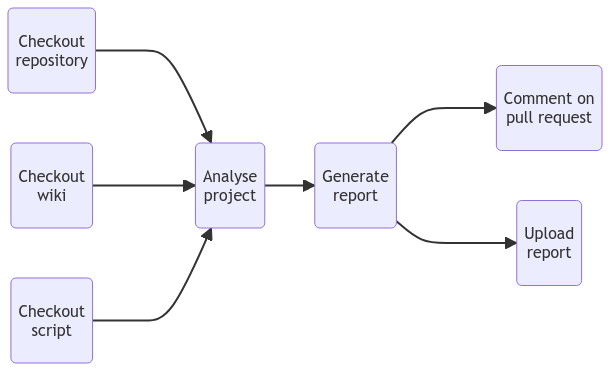}
    \caption{Overview of the steps automated using the tool}
    \label{fig:tool_overview}
\end{figure}

The GitHub repository, which includes the workflow outlined above and the source code for the \tool{} tool, is available for public access.\footnote{\url{https://github.com/wesleytanws/DOCER_tool/tree/v1.0.1}} \Cref{fig:tool_overview} summarises the steps defined by the workflow.

\subsection{Adding to GitHub projects}
\label{sec:tool_integration}
\begin{figure}[htbp]
    \centering
    \includegraphics[width=0.45\textwidth]{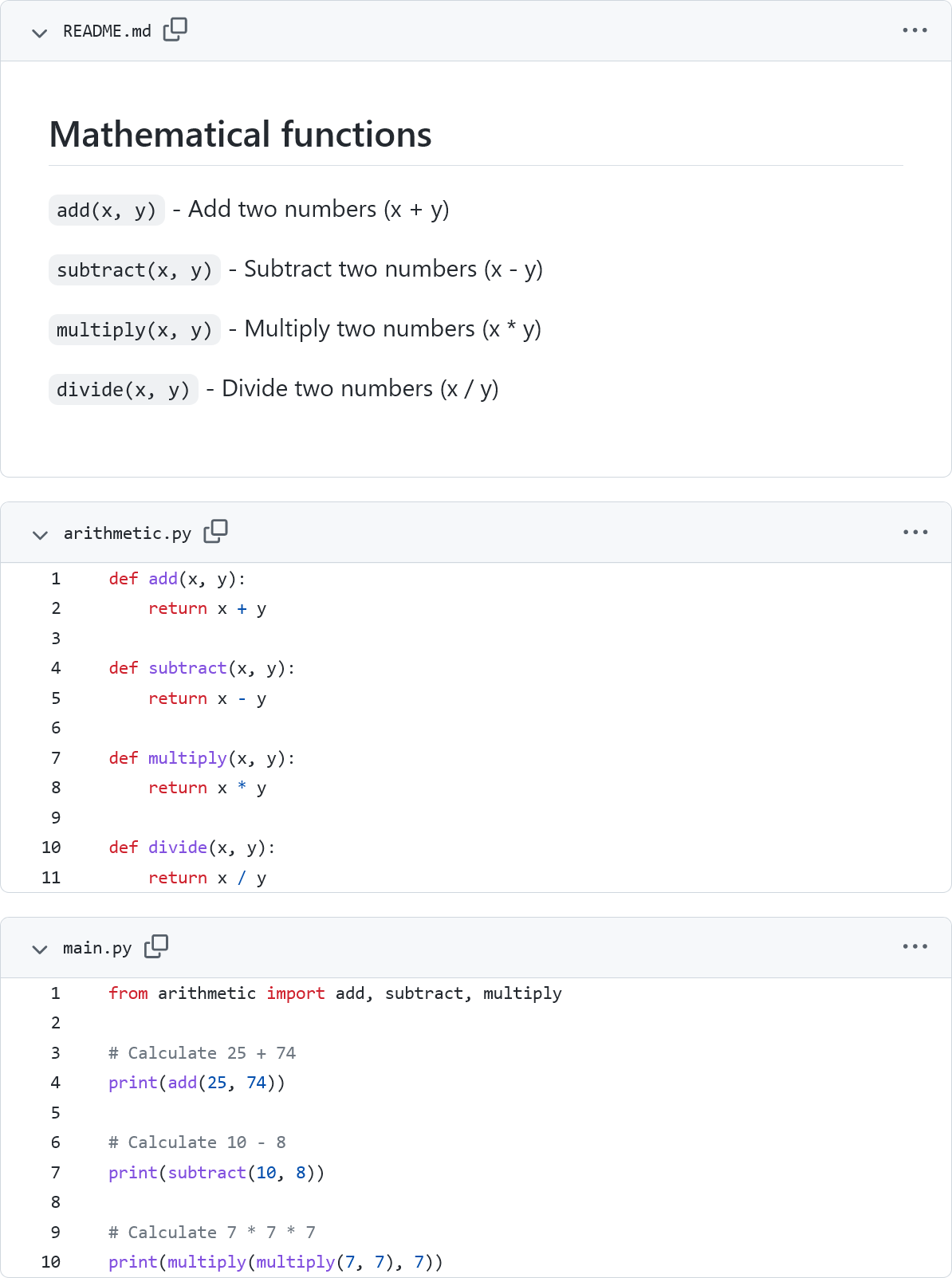}
    \caption{Files in the example repository for tool demonstration}
    \label{fig:tool_example}
\end{figure}

To demonstrate how the GitHub Actions tool works, we will integrate the tool with an example repository with three files (\Cref{fig:tool_example}):
\begin{itemize}
    \item \codeword{README.md} documents the mathematical functions defined in arithmetic.py
    \item \codeword{arithmetic.py} defines the mathematical functions
    \item \codeword{main.py} calls the functions defined in arithmetic.py
\end{itemize}

\begin{figure}[htbp]
    \centering
    \includegraphics[width=0.45\textwidth]{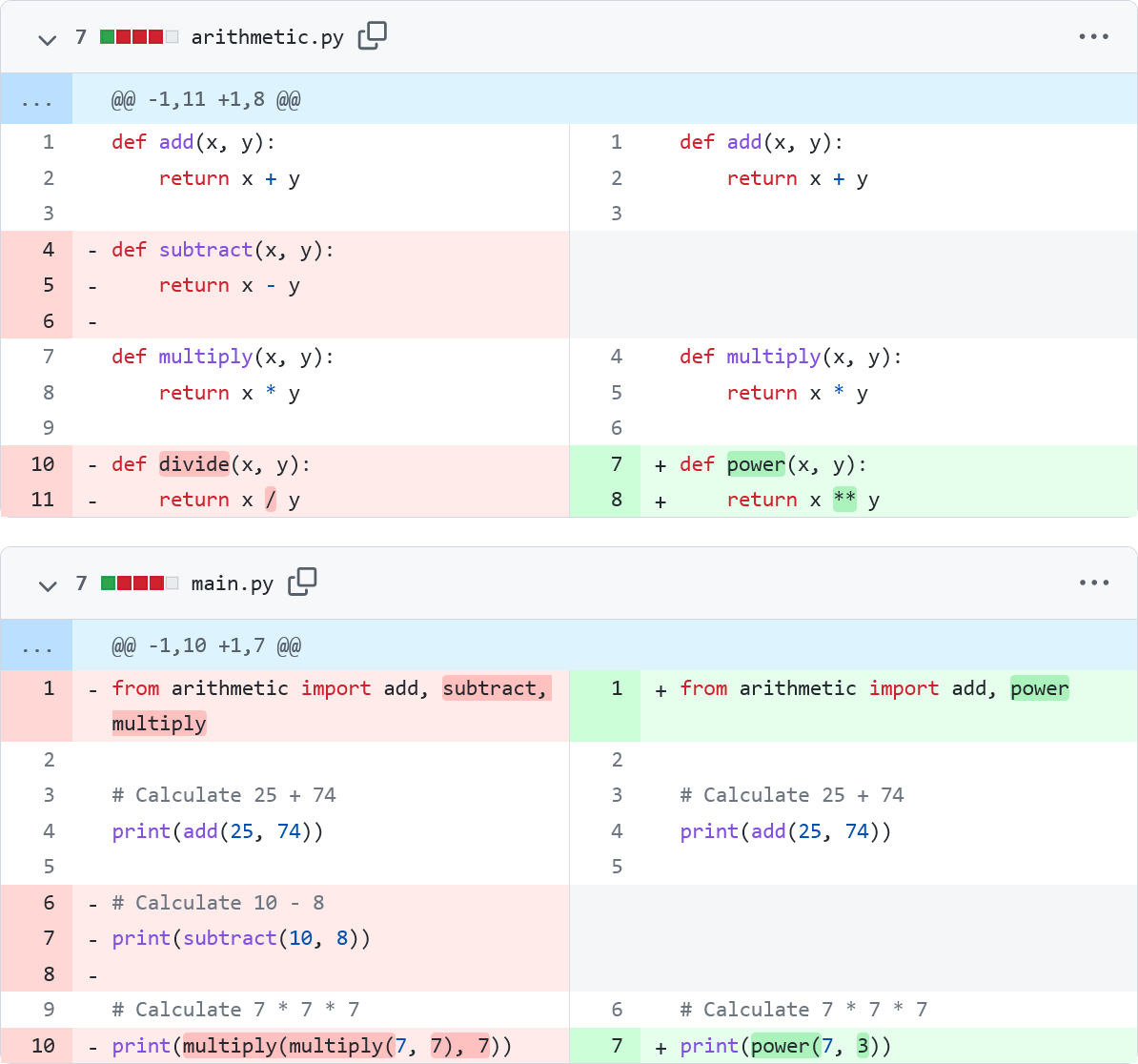}
    \caption{Pull request showing the incoming changes}
    \label{fig:tool_pr}
\end{figure}

Integrating the tool to a repository is as convenient as copying the YAML file defining the workflow\footnote{\url{https://github.com/wesleytanws/DOCER_tool/blob/v1.0.1/DOCER.yml}} to the \codeword{.github/workflows} folder. Suppose a pull request as shown in \Cref{fig:tool_pr} is submitted to the repository.

\begin{figure}[htbp]
    \centering
    \includegraphics[width=0.45\textwidth]{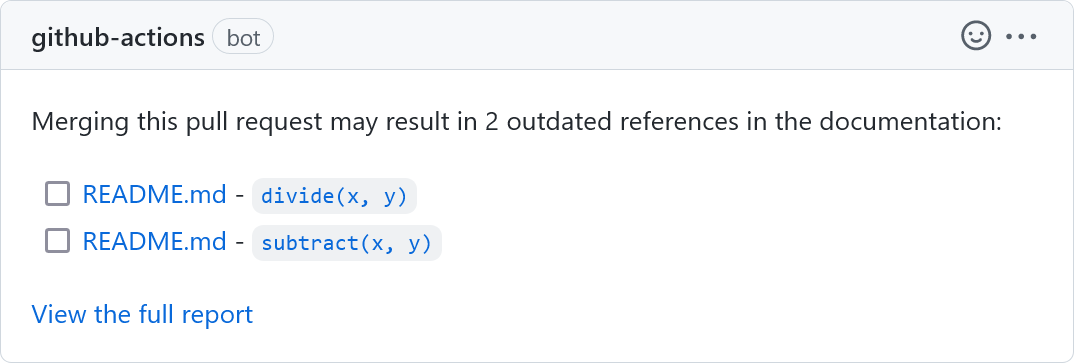}
    \caption{Comment on the pull request listing the potentially outdated code element references}
    \label{fig:tool_message}
\end{figure}

Looking at the pull request submitted, two files in the repository have been modified. In arithmetic.py, the subtract and divide functions were removed and a new power function was added. Similarly, the main.py file was modified to remove the subtract function and the chained multiply functions were refactored into a power function. Notice that the tool reports that continuing to merge the pull request may result in two outdated references in the documentation (\Cref{fig:tool_message}). This discrepancy arises because the README file was not updated to reflect the removal of `divide' and `subtract' functions from the source code.

\begin{figure}[htbp]
    \centering
    \includegraphics[width=0.45\textwidth]{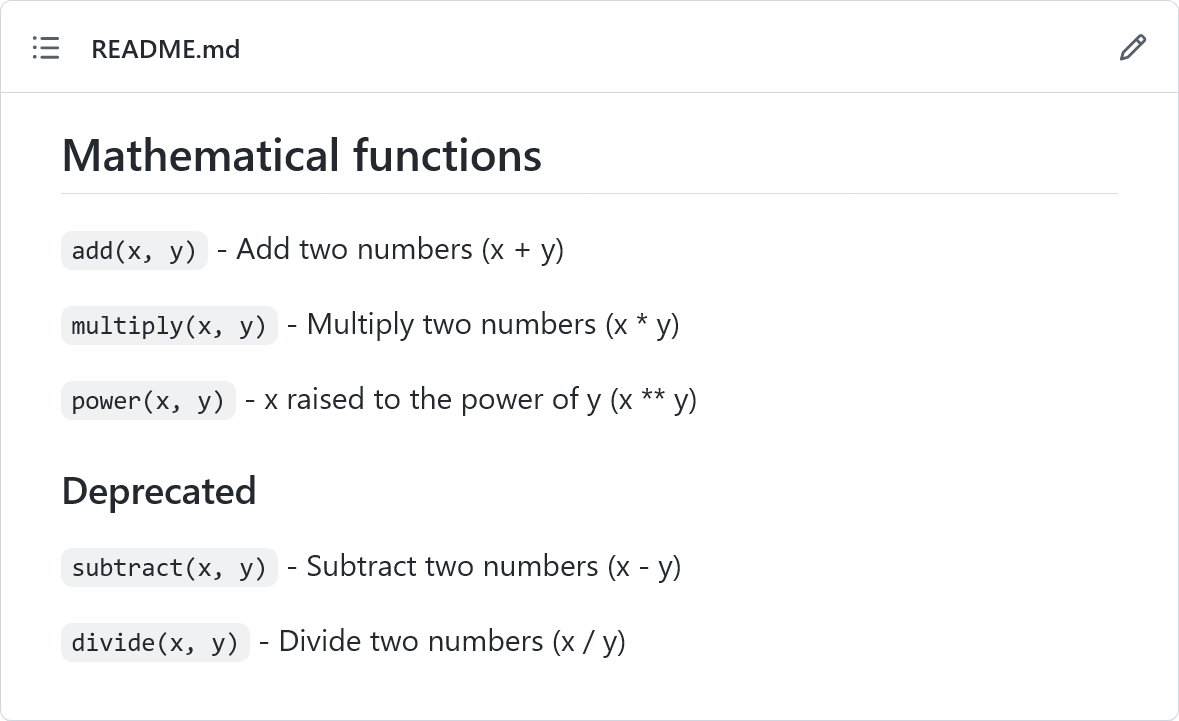}
    \caption{Updated README file including the new power function and listing the deleted functions as deprecated}
    \label{fig:tool_doc_updated}
\end{figure}

To keep the documentation up-to-date, we can simply remove the two outdated references in the README file. Better still, we can document the new function and mention that the two functions are now deprecated as shown in \Cref{fig:tool_doc_updated}.

\subsection{Excluding code elements}
\label{sec:tool_feature}
One useful feature that we added to the tool is the ability to exclude certain code elements from the report, which allows developers to stop keeping track of code elements that have been determined to be false positives. Developers can add a list of code elements separated by newlines in a file named \codeword{.DOCER\_exclude} located at the root of the repository. Code elements in the exclude list will be ignored by the tool when scanning for outdated references.

\section{Examples}
\label{sec:examples}
In our previous work~\cite{tan2022detecting}, we evaluated the approach's usefulness in real-world software projects by submitting GitHub issues to 15 different projects. Here, we present two examples of true positives and false positives in the issues submitted~\cite{tan2022detecting}. \tool{} automates the creation of such notifications.

\begin{figure}[htbp]
    \centering
    \includegraphics[width=0.49\textwidth]{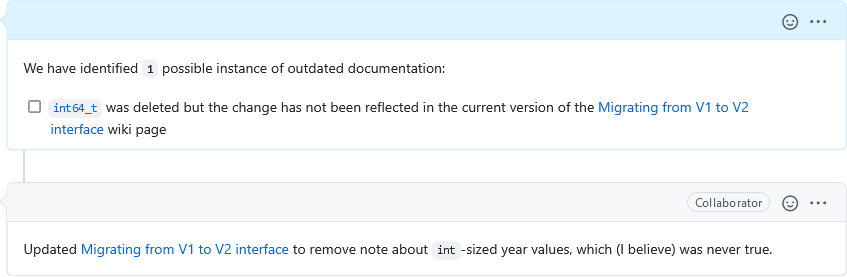}
    \caption{True positive: data type updated in the documentation}
    \label{fig:cctz}
\end{figure}

\begin{figure}[htbp]
    \centering
    \includegraphics[width=0.49\textwidth]{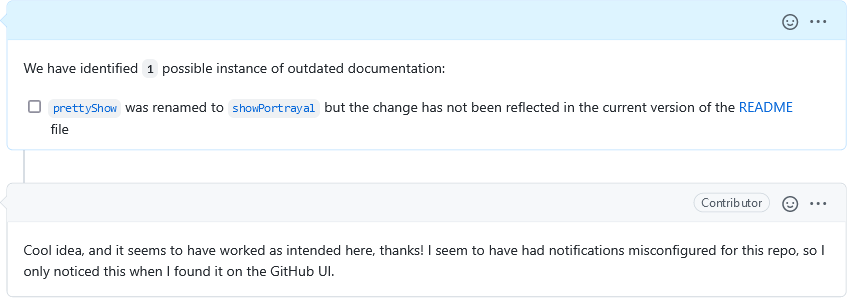}
    \caption{True positive: function name updated in the documentation}
    \label{fig:hs-portray}
\end{figure}

\paragraph{True positives} The google/cctz project was one of the 15 projects that responded positively to our GitHub issue.\footnote{\url{https://github.com/google/cctz/issues/210}} All code element instances \codeword{int64\_t} were removed from the source code in one of the commits but the documentation continued to reference the deleted code element. In response to our GitHub issue, the developer updated the documentation to align with the changes in the source code (\Cref{fig:cctz}). In the google/hs-portray project, the function \codeword{prettyShow} was renamed to \codeword{showPortrayal} in the source code, but the README file was not updated (\Cref{fig:hs-portray}). We alerted the developers of this discrepancy, and the issue was fixed subsequently.\footnote{\url{https://github.com/google/hs-portray/issues/7}}

\begin{figure}[htbp]
    \centering
    \includegraphics[width=0.45\textwidth]{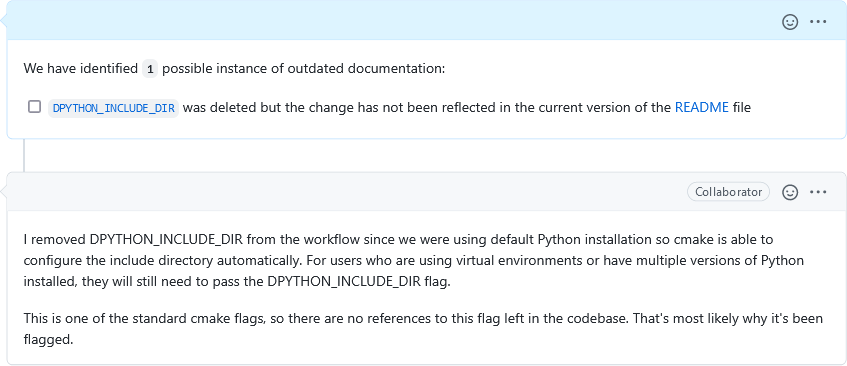}
    \caption{False positive: still relevant for users with multiple Python versions}
    \label{fig:clif}
\end{figure}

\begin{figure}[htbp]
    \centering
    \includegraphics[width=0.45\textwidth]{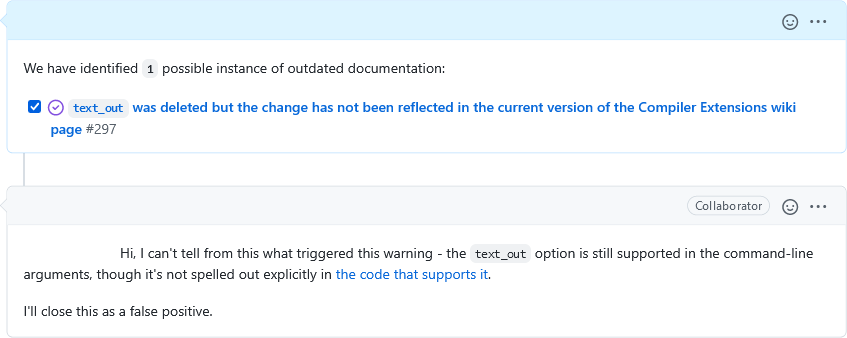}
    \caption{False positive: functionality remains in the program logic}
    \label{fig:gnostic}
\end{figure}

\paragraph{False positives} In another Google project google/clif (\Cref{fig:clif}), a CMake flag was removed from the source code but the documentation was not updated. The developer responded that the flag is no longer required in the source code but it is still relevant for users that have installed multiple versions of Python to configure the installation directory correctly.\footnote{\url{https://github.com/google/clif/issues/52}} A false positive was reported in the google/gnostic project (\Cref{fig:gnostic}) where the code element \codeword{text\_out} was deleted from the source code. Although the code element is no longer found in the source code, the functionality remains in the program logic. This leads to the code element reference getting falsely flagged as outdated.\footnote{\url{https://github.com/google/gnostic/issues/273}}

\section{Limitations}
\label{sec:limitations}
Trying to understand and use documentation which features code elements that do not exist is just one of many frustrations that software developers encounter when they are confronted with outdated documentation. Addressing this particular frustration is the goal of \tool{}. Other forms of outdated documentation, such as inaccurate descriptions of the functionality of code elements or not-yet-documented code elements, are beyond the scope of our current work. \tool{} is currently limited to detecting outdated documentation in GitHub (README and wiki pages) and would not be able to find issues in documentation hosted externally. \tool{} detects code elements in documentation using a set of regular expressions from previous work. These regular expressions have not been validated on all possible programming languages and refining them to work on popular programming languages is part of our future work.

Our tool may sometimes falsely categorised references as outdated due to limitations of the approach. For example, the change log of a project may contain references to deleted code elements in the source code. However, these references should not be flagged as outdated as they only serve as a notice. As a workaround, developers can add the code elements to the \codeword{.DOCER\_exclude} file to avoid the tool reporting the references as outdated. In addition, our tool only detects code elements written as text. Other kinds of outdated documentation such as images and videos in the documentation cannot be detected.

\section{Related Work}
\label{sec:related}

There are numerous existing work related to detecting and fixing inconsistencies between source code and documentation, with source code comments being one of the main focuses. Wen et al.~\cite{wen2019large} conducted an empirical study of 1500 Java systems, citing deprecation and refactoring as causes of code-comment inconsistencies. In one of the earliest attempts to address these inconsistencies, Tan et al.~\cite{tan2012tcomment} proposed @tcomment, aiming to catch exceptions related to null values in Javadoc comments. Ratol and Robillard~\cite{ratol2017detecting} introduced Fraco, a tool targeting source code comments and identifiers renaming. Panthaplackel et al.~\cite{panthaplackel2020learning} proposed a model that can modify natural language comments based on source code changes, outperforming existing comment generation models.

Other work related to documentation but not limited to source code comments include \textsc{DocRef} by Zhong and Su~\cite{zhong2013detecting}. Combining natural language tools and code analysis techniques to identify discrepancies between source code and documentation, \textsc{DocRef} was able to detect more than 1000 errors in API documentation. Designed to report documentation changes, AdDoc by Dagenais and Robillard~\cite{dagenais2014using} uses traceability links to identify changes to the documentation that deviate from existing code patterns. Using static program analysis, Zhou et al.~\cite{zhou2018automatic} proposed a framework DRONE, that automatically discovers defects in Java API documentation and generates helpful recommendations. Another work addressing API documentation is FreshDoc by Lee at al.~\cite{lee2019automatic}. By using a grammar parser and analysing multiple source code versions, FreshDoc can automatically update class, method and field names found in the documentation.

In contrast to these approaches and to the best of our knowledge, \tool{} is the first tool which attempts to prevent inconsistent and outdated documentation by alerting software developers before their documentation becomes outdated. We accomplish this through a GitHub Action which is GitHub's implementation of a software bot~\cite{lebeuf2017software}. Software bots have recently attracted the attention of the software engineering research community, with a particular focus on code review bots which---similar to \tool{}---comment on pull requests. For example, Wessel et al.~\cite{wessel2020effects} found that the adoption of code review bots increases the number of monthly merged pull requests, decreases monthly non-merged pull requests, and decreases unnecessary communication among developers. Our goal with \tool{} is to enable code review bots to also decrease the amount of outdated documentation.

\section{Future Work and Conclusion}
\label{sec:conclusion}
In this paper, we presented \tool{} that developers can use to automatically scan for outdated code element references. The tool analyses the repository and generates a report on the state of code element references whenever a pull request is submitted. If merging the pull request results in outdated references in the documentation, the tool will upload the report and comment on the pull request alerting developers of the situation. Developers can choose to fix the outdated references in their documentation, or add the references to the exclude list if they have been determined to be false positives.

As mentioned in \Cref{sec:limitations}, refining the list of regular expressions used to detect code elements is part of our future work. One such refinement could be ensuring that the regular expressions can accurately extract code elements found in popular programming languages such as JavaScript, Python and Java. In addition, several improvements can be made to the tool. Adding a feature where developers can reply to the tool's comment for code elements they do not want to keep track of could be helpful. The tool will then automatically add the code elements to the project's exclude list. Another improvement could be adding a file that defines a list of documentation files to exclude, e.g. wiki page that contains the project's change log. Expanding the tool to not only work on GitHub, but also other version control platforms is another direction worth exploring. This allows more developers to scan for outdated code element references in their projects.


\bibliographystyle{IEEEtran}
\bibliography{main}

\end{document}